\documentclass[%
 aip,
 jmp,%
 amsmath,amssymb,
 reprint,%
unsortedaddress
]{revtex4-1}

\usepackage{graphicx}
\usepackage{dcolumn}
\usepackage{bm}
\usepackage{float}
\usepackage{gensymb}
\usepackage{color}

\begin{document}
	\title{Photoinduced Charge Transfer in Transition Metal Dichalcogenide Quantum Dots}

		\author{Praveen Mishra} 
		\affiliation{School of Computational \& Integrative Sciences, Jawaharlal Nehru University, New Delhi-110067, India}
				\author{Arun Singh Patel} 
				\affiliation{Department of Physics, Dyal Singh College, University of Delhi, Lodhi Road, New Delhi – 110003, India}
				\author{Sanjay Kumar Chauhan} 
				\affiliation{Department of Physics, Hindu College, University of Delhi,  Delhi – 110007, India}
				
	\author{Anirban Chakraborti} 	
	\affiliation{School of Computational \& Integrative Sciences, Jawaharlal Nehru University, New Delhi-110067, India}
	\email{anirban@jnu.ac.in}


	\begin{abstract}
		In this paper, we have explored the charge transfer mechanism in transition metal dichalcogenide (TMDC) quantum dots (QDs) of molybdenum disulfide ($\rm{MoS_2}$) and tungsten disulfide ($\rm{WS_2}$). Rhodamine 6G (R6G), a dye from the rhodamine family, has been employed as the fluorescent molecule, with MoS$_2$ and WS$_2$ QDs acting as electron acceptors in the photo-induced charge transfer process. The TMDC QDs were synthesized using a top-down approach and characterized through transmission electron microscopy (TEM), UV-Vis spectrophotometry, and fluorimetry. TEM images revealed well-dispersed particles measuring 2 nm in size. These QDs exhibit strong fluorescence emission when excited with light at wavelengths below 350 nm. Under light exposure, photons generate charges in the fluorescent dye molecules, and the TMDC QDs facilitate the charge transfer process. The charge transfer phenomenon was investigated using time-correlated single photon counting (TCSPC), a time-resolved fluorescence spectroscopic technique. The time-resolved fluorescence study indicated a change in the fluorescence (FL) lifetime of R6G molecules in the presence of QDs. The FL lifetime of R6G molecules without QDs was found to be 4.0 ns, which decreased to 1.9 ns and 3.8 ns in the presence of MoS$_2$ and WS$_2$ QDs, respectively. This reduction in FL lifetime suggests that the MoS$_2$ and WS$_2$ QDs provide an additional pathway for photo-generated electrons in the excited state of R6G molecules. This research can be extended to optoelectronic devices, where charge transfer is crucial for device efficiency and performance.

	\end{abstract}
	
	
	\maketitle
In recent years, different kinds of quantum dots (QDs) have been extensively researched because of their numerous applications in cell imaging, photovoltaic devices, light-emitting devices, and more. \cite{xu2021quantum,li2024synergistic, wang2020full, sadeghi2019efficient, li2016cspbx3, xue2019toward, meng2024colloidal} The most appealing characteristic of these materials is their size-dependent opto-electronic properties, making them ideal candidates for designing various opto-electronic devices.\cite{li2015carbon, litvin2017colloidal}	The fascinating properties of QDs have been investigated in composite systems or devices through their combination with various materials, including molecules, other QDs, and 2D nanomaterials. When QDs are paired with biomolecules, they are utilized for bio-imaging, while QDs in a core-shell configuration are explored for optoelectronic devices. These QDs, when sensitized with other semiconductors, have been used to create cost-effective, next-generation photovoltaic devices for various applications. QDs are favored in these applications due to their size-dependent optical and electronic properties. The interaction between QDs and their surrounding media or materials is critical for device performance. Although the electronic interactions between organic molecules and fluorescent QDs have been well studied, systems where QDs are coupled with other inorganic molecules have been less explored. These interactions differ from those in QD-molecular systems because inorganic materials have a continuum of electronic states, whereas molecular acceptors have discrete electronic states. When QDs are coupled with other inorganic materials, they can donate electrons, as observed in photovoltaic devices, or act as electron donors/acceptors, as seen in light-emitting diodes (LEDs). In these applications, electron transfer involves a specific direction of transfer. Therefore, understanding the factors driving electron transfer in these composite systems is crucial for better comprehending and further exploiting the unique properties of QDs in optoelectronic devices.\cite{algar2008quantum, wang2021thermodynamic, chakraborti2016resonance, zhu2018planar, lima2022mannose, wu2020development, rani2020visible} 
	
	There are various studies on different kind of QDs, where these QDs have been used as acceptor/donor or both of the electrons. They have also been  explored for donor or acceptor in the resonance energy transfer. In recent years researchers have explored various kinds of layered based QDs in place of traditional QDs, e.g., CdS, CdSe, PbS, etc.\cite{kharangarh2023synthesis, sariga2023new, gan2015quantum, guo2019ws2} In optoelectronic applications, QDs function as light-harvesting and charge-separation centers. A crucial step in this process is the transfer of electrons or holes from the photoexcited QDs to nearby acceptors through a nanoscale interface. The overall efficiency of the device relies on these charge transfer processes, which are influenced by competing loss pathways. Therefore, understanding and controlling interfacial charge transfer processes is vital for improving QD-based devices. The size of the QDs also significantly impacts the charge transfer process, with studies showing that the electron transfer rate increases as the QD size decreases. \cite{zhu2016charge}

In recent years, various 2D layered nanomaterials have been transformed into QDs, which exhibit significantly different properties compared to their 2D structures and bulk counterparts. Specifically, 2D layered transition metal dichalcogenides (TMDCs) have been converted into QDs using various methods. Among these TMDC-based QDs, MoS$_2$ and WS$_2$ QDs have been extensively studied. They exhibit intriguing optical and electronic properties, with their electronic structure being tunable by adjusting their size and surface morphology. These QDs have been applied in various fields, including sensors, LEDs, photodetectors, and cell imaging. \cite{yan2016facile, ghorai2017highly, singh2019ws2, guo2020mos2}  
	
Several studies have used TMDC QDs as energy donors in hybrid systems due to their photoluminescence in the UV region. There is extensive research on energy transfer and charge transfer in various nanomaterials and their applications in sensors, photodetectors, and solar cells. In this work, we have investigated charge transfer in two different types of fluorescent materials: fluorescent dye molecules and QDs. Typically, QDs are used as donors or acceptors, and dye molecules as acceptors or donors in energy transfer studies, which often focus on the energy transfer and its applications. Here, we have explored the potential for charge transfer from dye molecules to QDs, providing valuable insights for developing optoelectronic devices based on fluorescent nanomaterials. In this paper, we used a common dye molecule, R6G, as the source of photo-excited electrons and TMDC QDs as the electron acceptors. 
	
	In order to synthesize the QDs and perform this study, various chemicals were used. Molybdenum disulphide (MoS$_2$), tungsten disulphide (WS$_2$)  powder (2$\mu$m size), N-Methyl 2-pyrrolidone (NMP), and Rhodamine 6G dye were purchased from Sigma Aldrich. All the chemicals were used as received without further purification.
		
\begin{figure}
		\centering
		\includegraphics[width=0.95\linewidth]{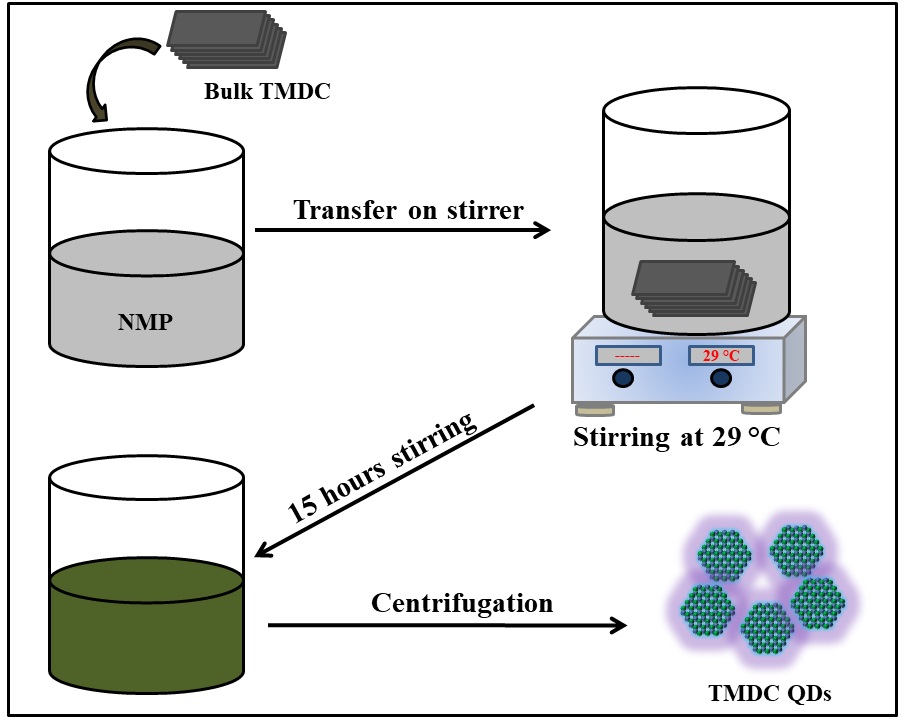} 
		\caption{Schematic diagram for the synthesis of TMDC quantum dots.      \label{Schematic}}
	\end{figure}

	The TMDC QDs were synthesized using top-down strategy as mentioned in Long et al. with slight modification.\cite{long2016ws2} In typical synthesis procedure,  50 mg of TMDC powder was dissolved in 50 mL of NMP solution with constant stirring for 15 hours at 29 $^\circ$C. The resulting dark suspension was centrifuged at 10,000 rpm for 40 min. The supernatant of the solution was collected and used for further studies.	The schematic diagram of the synthesis  of TMDC quantum dots is shown in Fig \ref{Schematic}.

	In our study, prior to exploring charge transfer mechanism,   the quantum dots (QDs) were characterized using transmission electron microscopy (TEM), fluorescence spectroscopy, and absorption spectroscopy. TEM images were captured with a JEOL-2010 TEM operating at 200 kV. The absorption spectra of the QDs were obtained using a Shimadzu (UV-2450) UV-Vis spectrophotometer.   In order to study charge transfer in these TMDCs QDS, the QDs were mixed with a fixed quantity of R6G solution. For charge transfer studies, 0.5 mL aqueous solution of 10$^{-6}$ M R6G was further diluted to 3.5 mL, and 1 mL of QDs solution  was mixed to R6G solution. In all the samples the NMP and water contents were kept same.  The steady-state fluorescence spectra of R6G molecules were captured both in the absence and presence of TMDCs QDs using a fluorimeter with an excitation wavelength of 430 nm. The time-resolved fluorescence analysis was conducted using a time-correlated single-photon counting module (TCSPC) (FL920, Edinburgh Instruments, UK), with excitation provided by a diode laser at 465 nm.

							\begin{figure}
								\centering
								\includegraphics[width=0.95\linewidth]{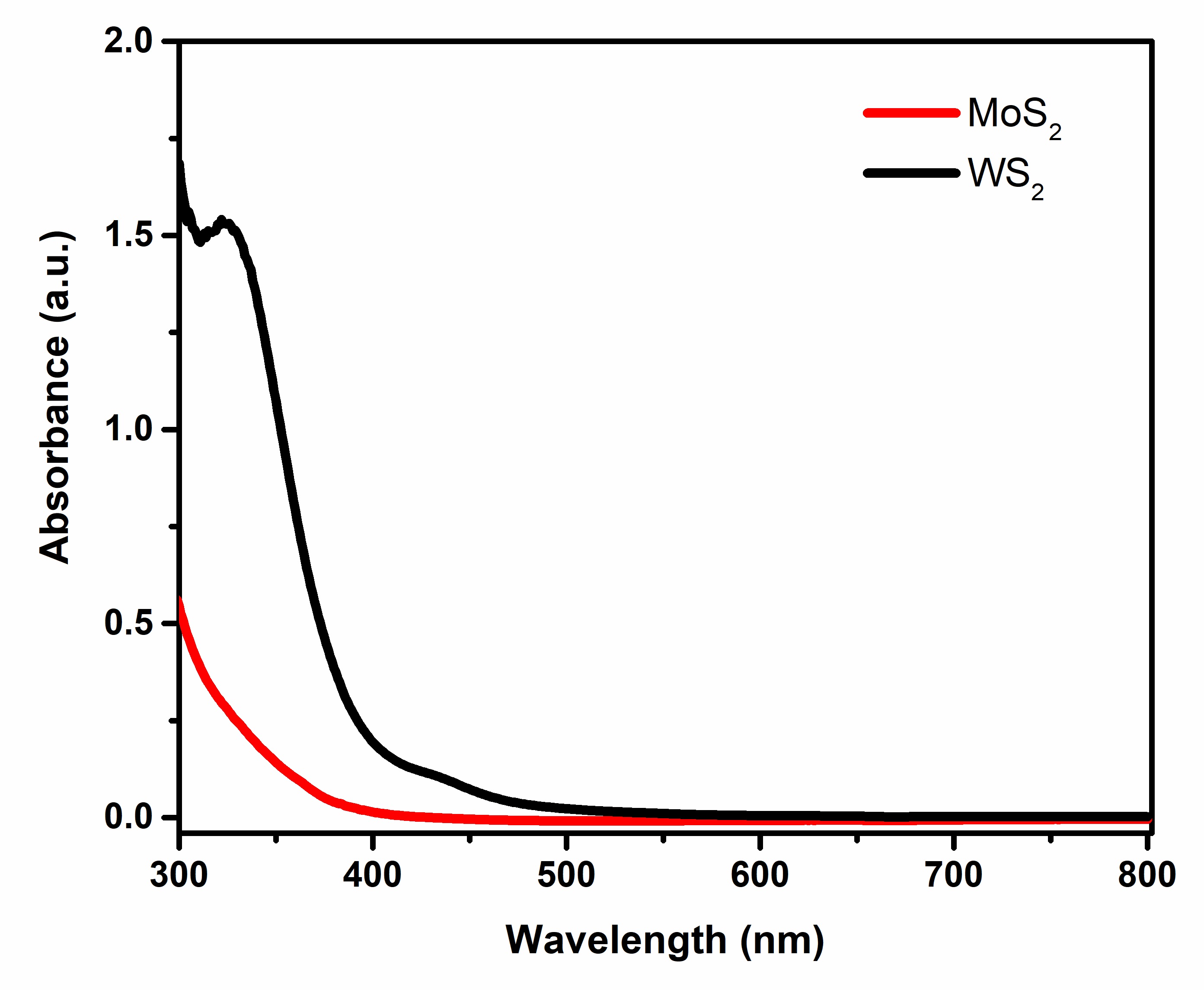} 
								\caption{Absorption spectra of MoS$_2$, and WS$_2$ quantum dots.     \label{Abs_MoS2_WS2}}
							\end{figure}
							
\begin{figure}
		 						\centering
		 						\includegraphics[width=0.95\linewidth]{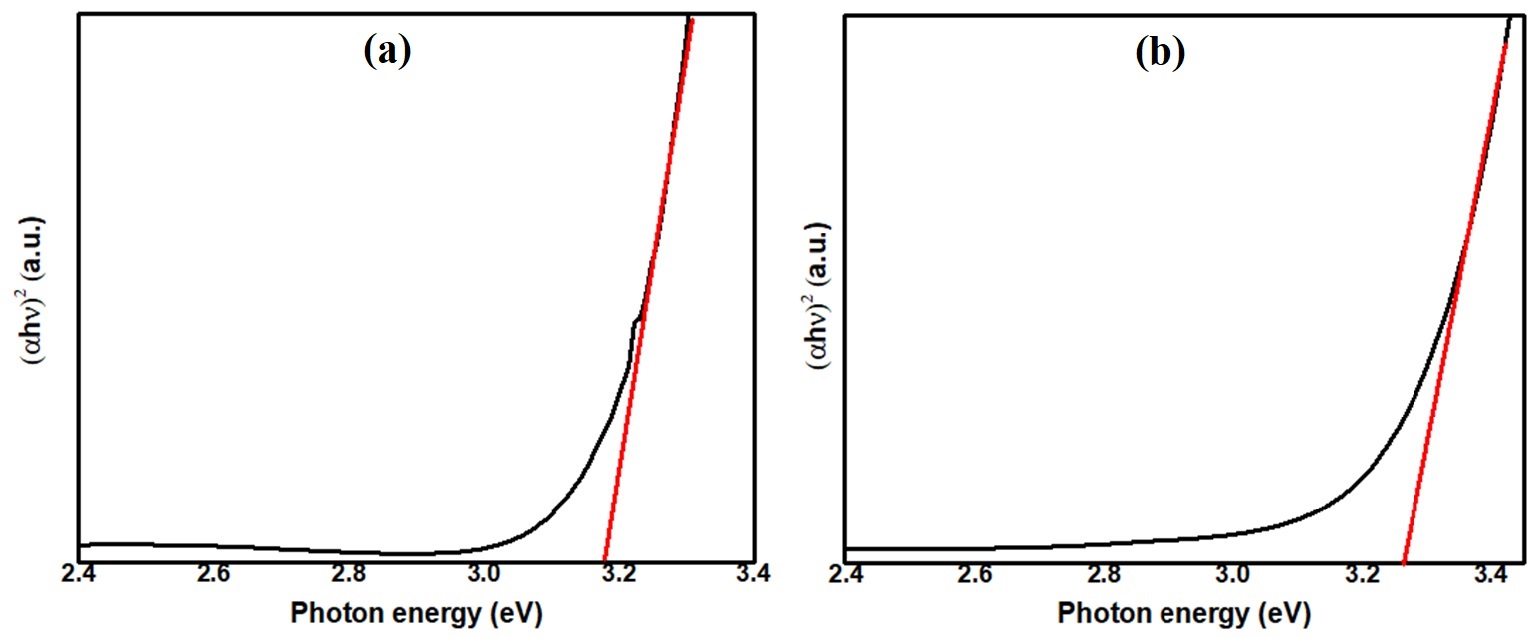} 
		 						\caption{The Tauc plots of (a) MoS$_2$, and (b) WS$_2$ quantum dots.     \label{Taucplot}}
		 					\end{figure}

 The absorption spectra  of MoS$_2$ and WS$_2$ quantum dots are shown in Fig \ref{Abs_MoS2_WS2}. A clear absorption peak around 325 nm is observed, which is a characteristic peak of  WS$_2$ quantum dots and it is due to excitonic features of the QDs. Similarly, for MoS$_2$ QDs a band edge around 380 nm is observed. The color of suspension and the absorption peaks are similar to the previous reported results. \cite{coleman2011two, long2016ws2}  When the MoS$_2$ and WS$_2$ are in the bulk or nanosheet form, they exhibit quite different absorption patterns having two distinct peaks around 600 nm which are known as $A$ and $B$ excitonic peaks. While in case of QDs these $A$ and $B$ excitonic peaks disappear and absorption is observed in UV region.  The pronounced blue shift observed in the absorption spectra of the synthesized MoS$_2$ and WS$_2$ QDs can be attributed to the quantum confinement effect and edge effects that become significant as the lateral dimensions of the QDs are reduced to a few tens of nanometers. \cite{li2017preparation, liu2014large}
		 In order to have insight of the band gaps of these QDs, the  Tauc plots of these QDs are drawn.  The  Tauc plots of the quantum dots are shown in Fig \ref{Taucplot}. 
		 
		 In case of semiconductors, it has been observed the bandgap increases with decrease in the size of the semiconducting quantum dots.  The size dependent bandgap is given by following equation:\cite{wu2005experimental}
		 \begin{equation}
		 E^*=E_g+\frac{h^2}{8\mu r^2}-\frac{1.8e^2}{4 \pi k\epsilon_0 r}\label{bandgap}.
		 \end{equation}
		  		
	In this expression, $E_g$ represents the bandgap of the bulk material, $h$ is Planck's constant, $\mu$ is the reduced mass of the exciton, $r$ is the quantum dot radius, $e$ is the electron charge, $\epsilon_0$ is the permittivity of free space, and $k$ is the dielectric constant of the material. This approximation assumes the particles are spherical, and the motions of the electron and hole are described by their effective masses. The effect of surface atoms are also ignored in this approximation. For MoS$_2$, the reduced mass of exciton is 0.27 times that of the free electron, the bulk bandgap $E_g$ is 1.29 eV,	and the permittivity  is about 6.8 times that of free space. The calculated value of the bandgap for MoS$_2$ quantum dots is found to be 1.9 eV. Similarly for WS$_2$ the bandgap was found to be 2.7 eV. In case of WS$_2$, the reduced mass was taken as 0.146 times  that of the free electron mass, the dielectric constant of the material was taken as 4.13 while theoretically calculation of the band gap. In the Equation \ref{bandgap}, the second and third terms of the equation are the kinetic energy and electrostatic  attraction terms, respectively. It is the  kinetic energy term which dominates in general, and results  a blue-shift in the bandgap as the size of the quantum dots decreases. As the size of QDs decreases, the bandgap increases because the valence band shifts to more positive potentials, while the conduction band shifts to more negative potentials. In most semiconducting QDs, this increase in bandgap is primarily due to the conduction band moving to more negative potentials, with only a slight shift in the valence band. (considering potential relative to NHE). \cite{gan2015quantum, wu2005experimental, parsapour1996electron} The calculated bandgap is found to be different from the experimentally observed values using Tauc plot which is due to the different size of the QDs, hence there might be variation in the reduced mass of the excitons and the variation in the dielectric constants values at the nanometer scale. 
	
					\begin{figure}
						\centering
						\includegraphics[width=0.95\linewidth]{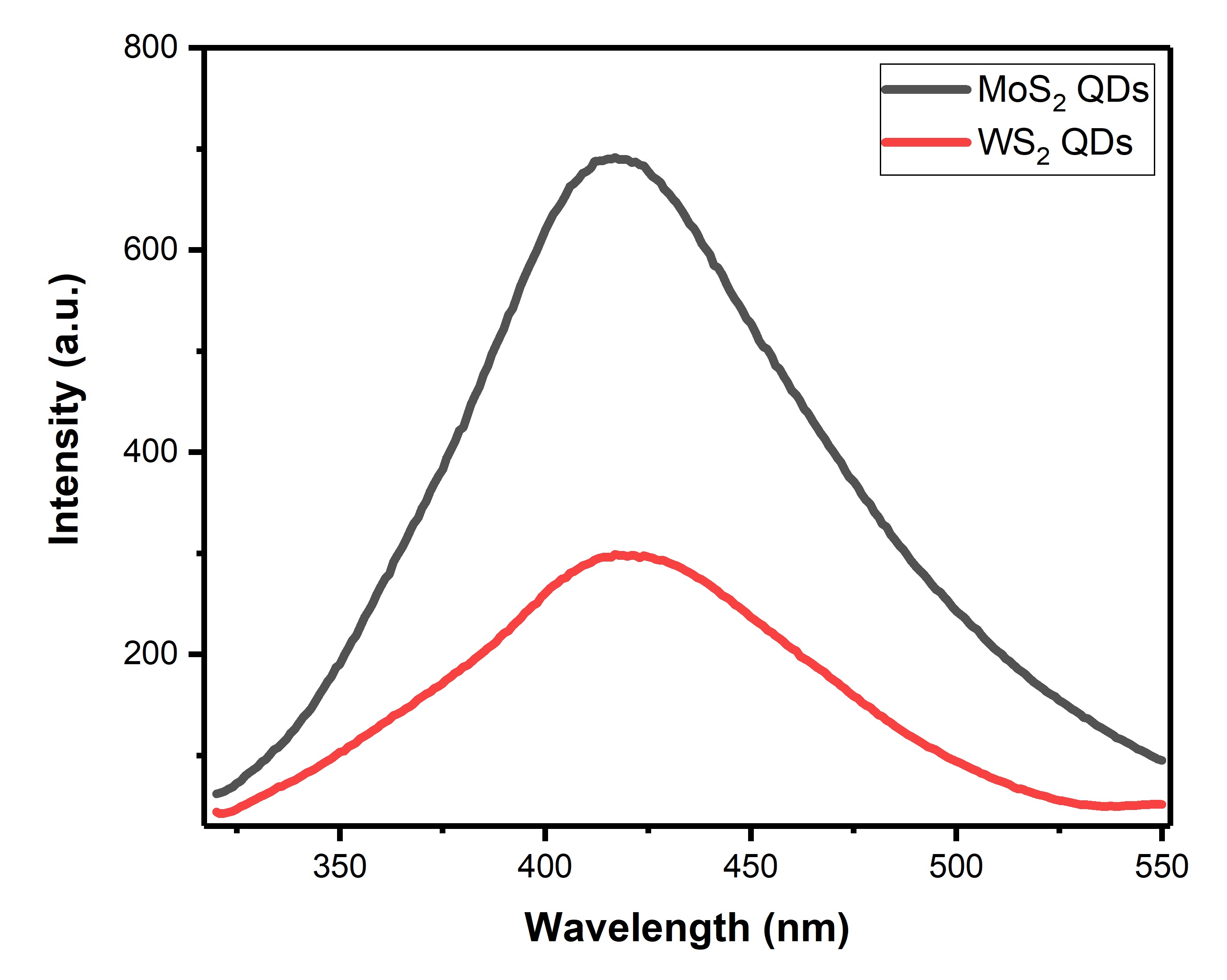} 
						\caption{Fluorescence spectra of MoS$_2$, and WS$_2$ quantum dots.     \label{QDsFL}}
					\end{figure}
		The fluorescence spectra of MoS$_2$, and WS$_2$ are shown in Fig. \ref{QDsFL}. The QDs show emission in blue region of the electromagnetic spectrum. The origin of this fluorescence emission is quantum confinement effect and the diract bandgap transition in these quantum dots. The emission spectrum has a wide pattern which is due to variation in the size of these QDs. The distribution of these particles size is having broad nature. \cite{ha2014dual}

			\begin{figure*}
				\centering
				\includegraphics[width=0.8\linewidth]{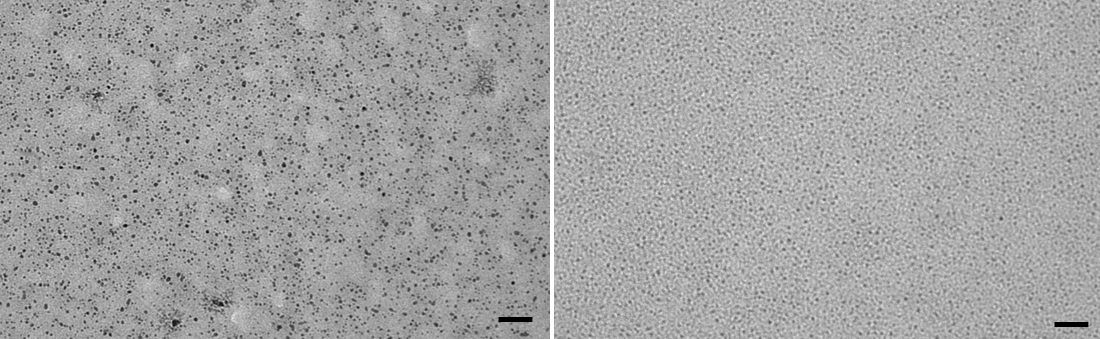} 
				\caption{TEM images of  (a) MoS$_2$, and (b) WS$_2$ quantum dots. Scale bar equals to 20 nm.     \label{TEM}}
			\end{figure*}
	The structural properties of  MoS$_2$ and WS$_2$ quantum dots were explored by using  transmission    electron microscopy (TEM). Fig. \ref {TEM} shows the TEM images of the quantum dots.  	The average size of QDs is of the order of 1.2 nm in case of WS$_2$ QDs, and 1.7 nm for the MoS$_2$ QDs.

\begin{figure}
			\centering
			\includegraphics[width=0.95\linewidth]{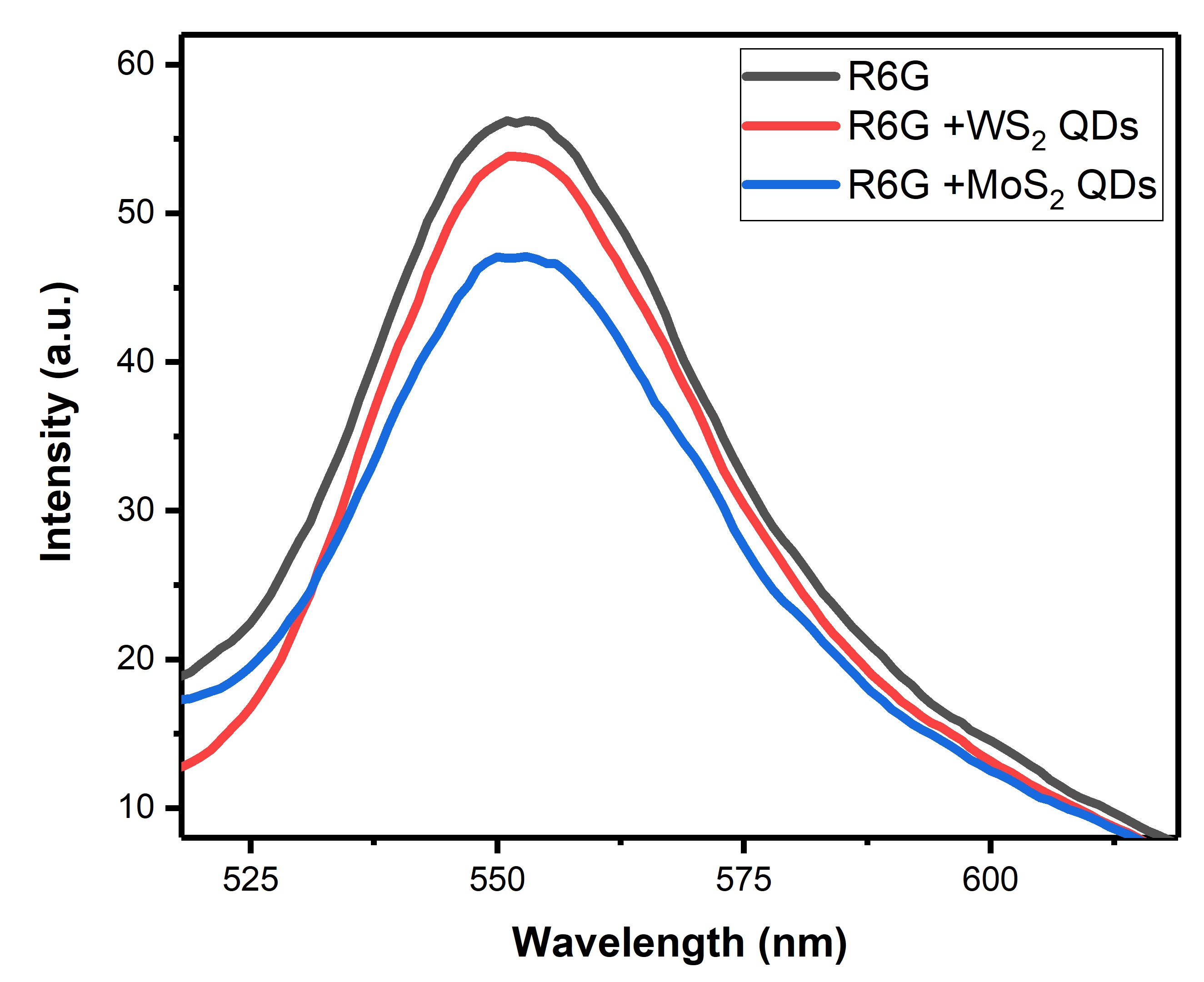} 
			\caption{Fluorescence spectra of R6G molecules in absence and presence of   (a) MoS$_2$, and (b) WS$_2$ QDs.   \label{figure_R6GFl}}
		\end{figure}

\begin{figure}
			\centering
			\includegraphics[width=0.95\linewidth]{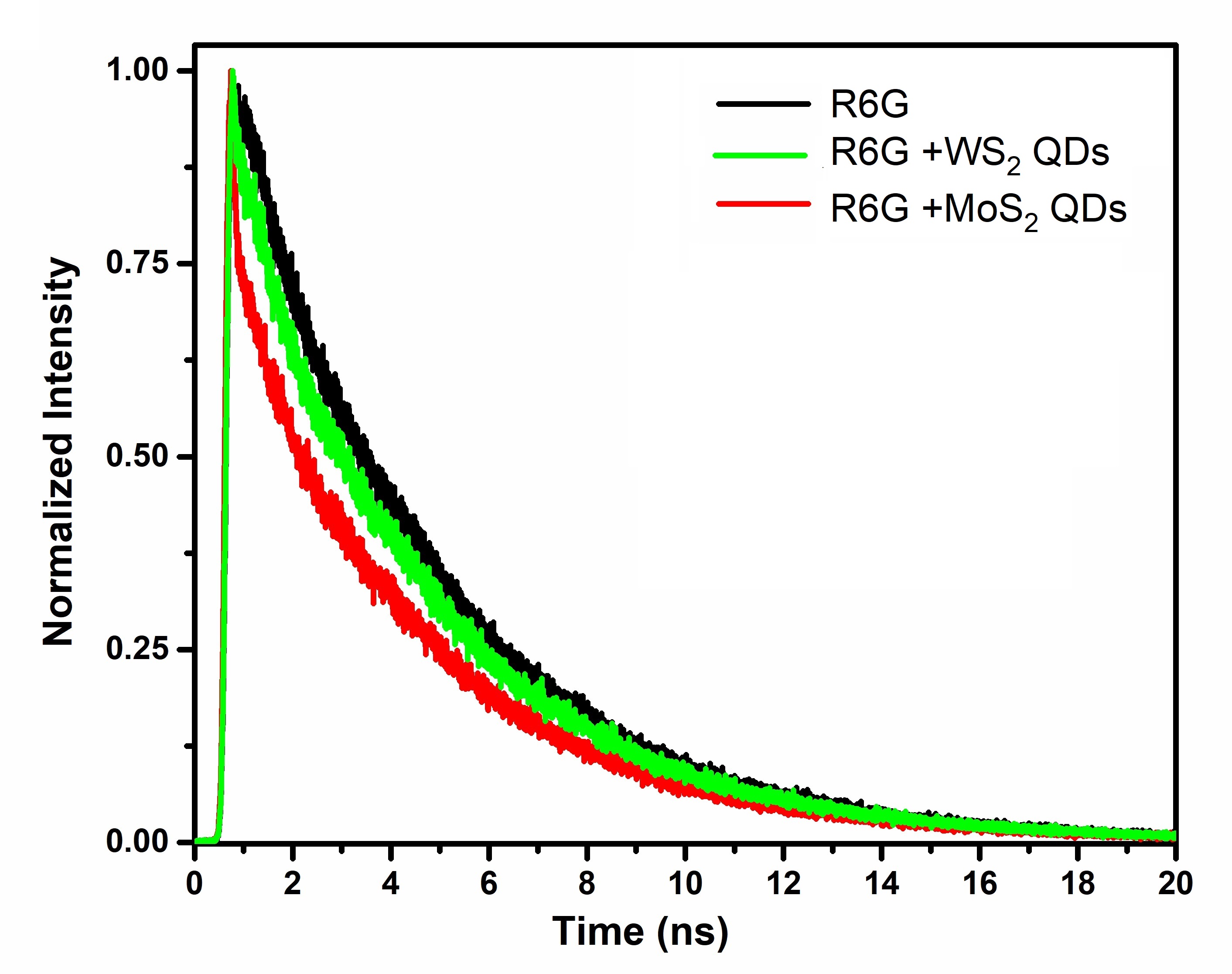} 
			\caption{Fluorescence lifetime decay curves of R6G molecules,  in the absence and  presence of MoS$_2$, and  WS$_2$ quantum dots.   \label{figure_TRFS}}
		\end{figure}

\begin{figure}
			\centering
			\includegraphics[width=0.95\linewidth]{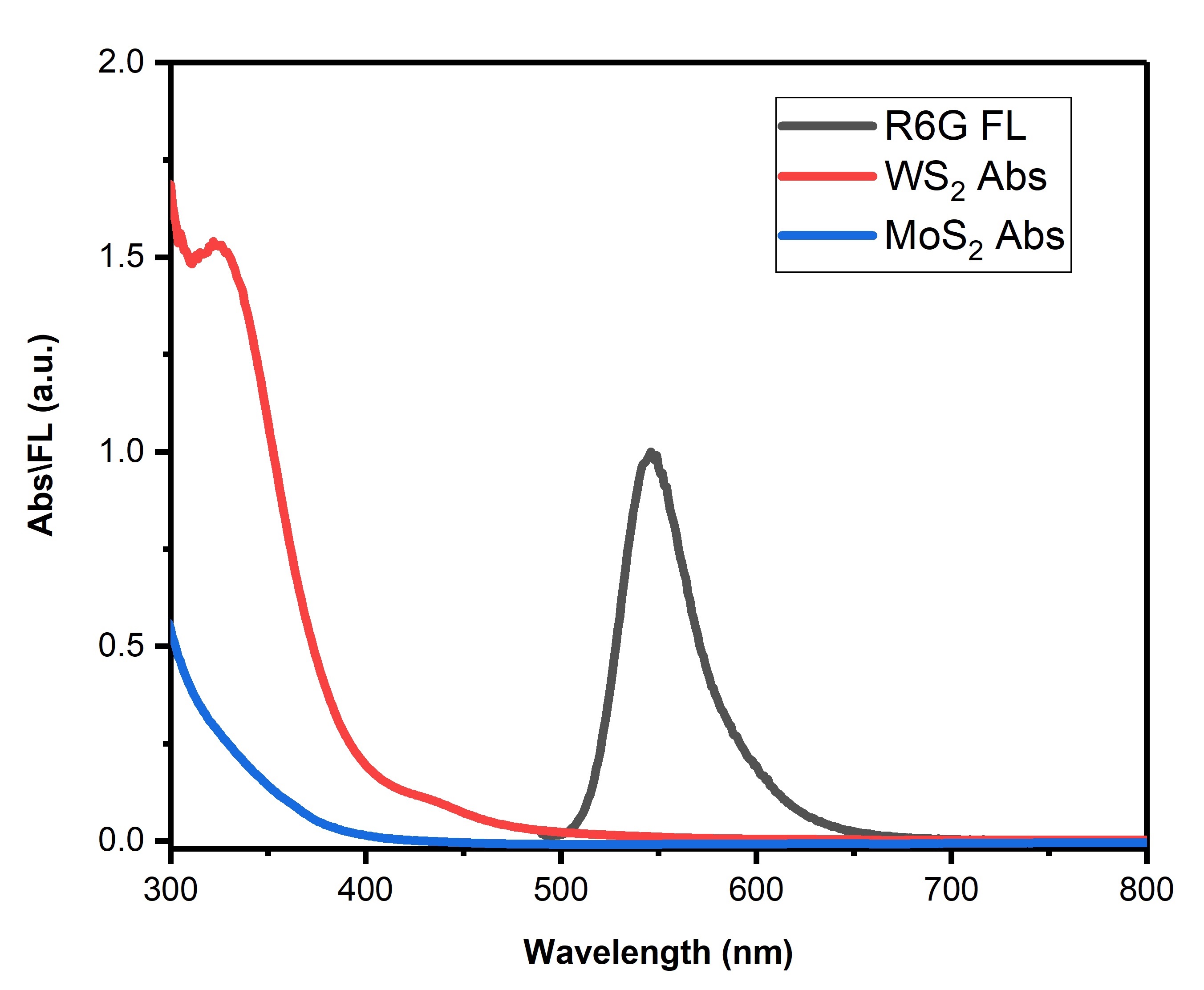} 
			\caption{Absorption spectra of   MoS$_2$, and  WS$_2$ QDs, along with the fluorescence spectrum of R6G molecules.   \label{figure_abs}}
		\end{figure}

The photo-response of R6G molecules, both in the absence and presence of QDs, was investigated using fluorescence spectroscopy. The steady-state fluorescence spectra of R6G molecules, with and without MoS$_2$ and WS$_2$ QDs, were recorded using a fluorimeter with a 430 nm excitation wavelength. Emission measurements were taken across a wavelength range of 520-620 nm. The steady state fluorescence spectra of R6G molecules in the absence and presence of MoS$_2$, and WS$_2$  are shown in Fig. \ref{figure_R6GFl}. The R6G molecules show characteristic emission peak  around 550 nm. 
In Fig. \ref{figure_R6GFl}, it was observed that the fluorescence  intensity of R6G decreases in the presence of QDs, indicating quenching by the QDs. This quenching could be either dynamic or static, which can be confirmed through time-resolved fluorescence studies. To investigate potential charge transfer from R6G to the QDs, the fluorescence lifetime of R6G molecules was measured both with and without the QDs using a time-correlated single photon counting (TCSPC) setup. The resulting fluorescence lifetime decay spectra of R6G in both scenarios are depicted in the Fig. \ref{figure_TRFS}. 
Fig. \ref{figure_TRFS} presents the fluorescence  lifetime decay curves of R6G molecules both in the presence of MoS$_2$ and WS$_2$ quantum dots (QDs) and in their absence. These decay curves were analyzed using exponential decay functions of the following form: \cite{patel2014}
 \begin{equation}
 	I(t)=\sum\limits_{j=1}^{m}\alpha_j \exp(-t/\tau_j), \label{eq_expfit}
 \end{equation}
The expression for the fluorescence decay includes $m$ as the number of discrete decay components, with $\tau_j$ representing the decay times, and $\alpha_j$ as the weighting factors corresponding to the $j^{th}$ decay component. It was observed that the fluorescence decay curves steepened in the presence of quantum dots (QDs) compared to pure R6G. While the R6G decay curve alone was well-fitted with a single-exponential function ($m=1$), in the presence of QDs, a bi-exponential function ($m=2$) was required for fitting. The average fluorescence lifetime for R6G alone was 4.0 ns, but this decreased to 1.9 ns with MoS$_2$ QDs and 3.8 ns with WS$_2$ QDs. The reduction in fluorescence lifetime suggests charge or energy transfer from R6G molecules to the QDs, with the effect being more pronounced for MoS$_2$ QDs. To further investigate the reasons for this decrease in fluorescence lifetime, the absorption spectra of the QDs were compared with the emission spectrum of R6G molecules, as depicted in Fig. \ref{figure_abs}.

In Fig. \ref{figure_abs}, the emission spectrum of R6G does not overlap with the absorption spectra of TMDC QDs, which rules out energy transfer from R6G to the QDs as a cause for the observed reduction in fluorescence intensity and lifetime. Instead, the decrease is likely due to charge transfer from photo-excited R6G molecules to the QDs. This charge transfer process leads to quenching of both fluorescence intensity and lifetime. The charge transfer efficiency $E$ from R6G to TMDC QDs was calculated using the following equation:
   \begin{equation}
   	E=1-\frac{\tau_{DA}}{\tau_D} .
   	\label{eq_energytime1}
   \end{equation}
Here, $\tau_{D}$ and $\tau_{DA}$ are fluorescence lifetimes of R6G molecules, in absence and presence of TMDCs QDs (acceptors), respectively.    The charge transfer efficiency, in case of MoS$_2$ QDs is found to be 53\%. For WS$_2$, the efficiency is found to be 3 \%.

\begin{figure}
	\centering
	\includegraphics[width=0.95\linewidth]{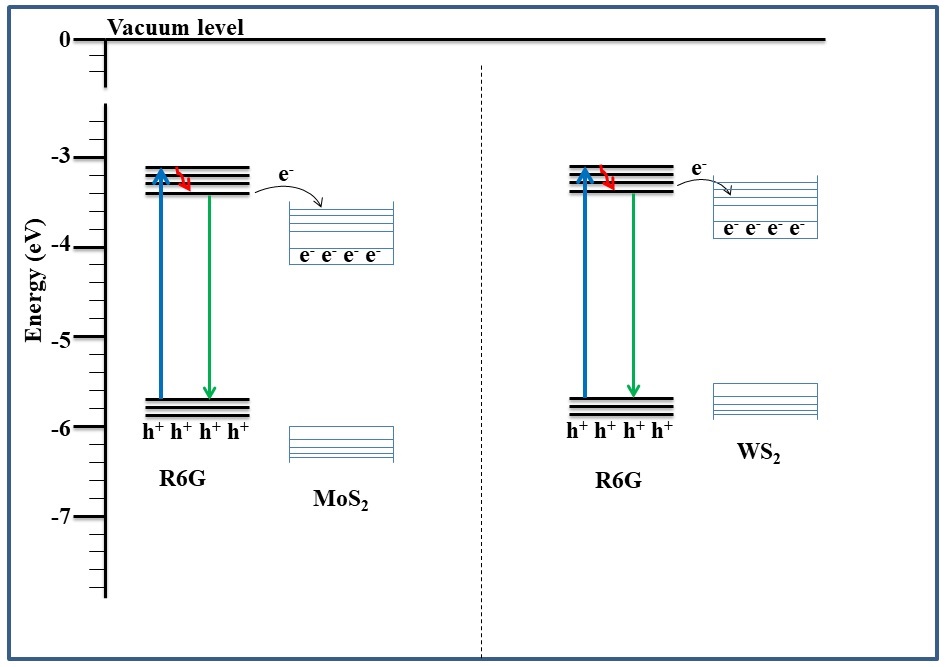} 
	\caption{Schematic of charge transfer from R6G molecules to MoS$_2$, WS$_2$ quantum dots. \label{CT_EL}}
\end{figure}

\begin{figure}
	\centering
	\includegraphics[width=0.95\linewidth]{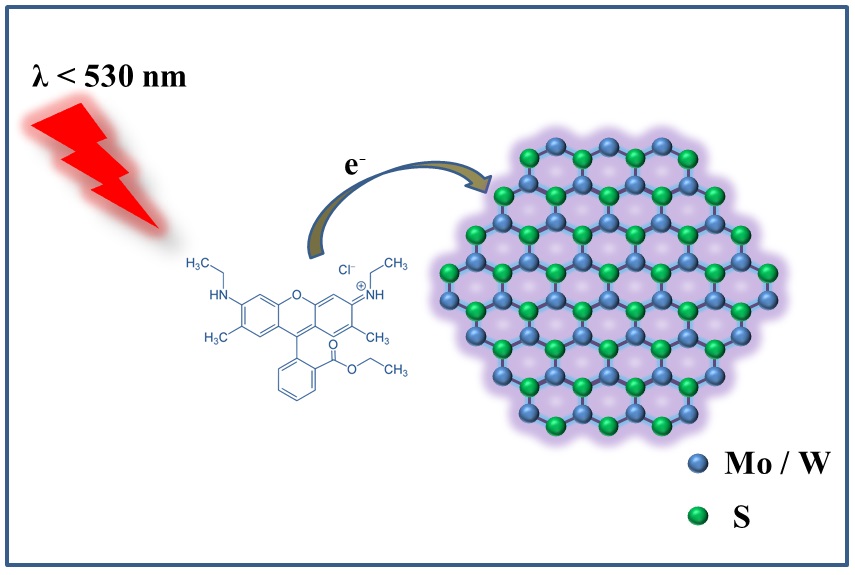} 
	\caption{Schematic of charge transfer from R6G molecules to MoS$_2$, WS$_2$ quantum dots. \label{Charge_trans}}
\end{figure}

The mechanism of charge transfer from R6G molecules to TMDCs QDs is illustrated in Fig.  \ref{CT_EL}, and Fig. \ref{Charge_trans}.   	The schematic of energy levels of R6G, MoS$_2$ and WS$_2$ quantum dots are   shown in Fig. \ref{CT_EL}, and a schematic of charge transfer in the QDs are shown in Fig. \ref{Charge_trans}.

It has been previously observed that the dye can be brought to excited state by absorbing the photons of suitable wavelength  and further the excited dye can transfer electrons to the conduction band  of  adjacent semiconducting materials, in this way the dye can be degraded by the reactive oxygen species.\cite{li2010comparison} Here, in this study, we used R6G dye molecules which has redox potentials 1.2 V and -1.1 V vs NHE for  ground state (HOMO) and excited state (LUMO), respectively.\cite{li2010comparison}  The energy levels  of R6G in its ground state and excited state versus vacuum is calculated using equation E(eV) = -4.5 -E$_{NHE}$ and the respective energy levels are -5.7 eV and -3.4 eV, for ground and excited state, respectively. The optical band gap of R6G dye molecules is of the order of 2.3 eV, which corresponds to 540 nm of the wavelength. The potential of the valence  band  and conduction band	 of MoS$_2$ are found to  be -6.0 eV and -4.2 eV, respectively. Similarly, the valence band and conduction band energy levels of WS$_2$ are found to be at -5.48 eV, and -3.93 eV, respectively. \cite{mukherjee2016novel} From these energy levels, it is observed that by absorbing photons of wavelength $\lambda$ $<$ 530 nm, R6G molecules make transition into an excited state R6G$^*$ and the energy differences between R6G* state to the conduction bands of QDs provide a thermodynamically favourable condition for the transfer of electrons from LUMO of R6G to the conduction band of MoS$_2$ and WS$_2$ QDs. The energy difference between LUMO of R6G and conduction band of MoS$_2$  is 0.8 eV and for   WS$_2$, the difference is 0.53 eV. The energy differnce is maximum in case of MoS$_2$ which favours the charge transfer from R6G molecules in the excited state to the conduction band of the MoS$_2$ as compared to the  WS$_2$. Therefore the charge transfer efficiency is maximum (53$\%$) in  MoS$_2$ QDs. This study can be further extended to other quantum dots and 2D materials and the efficiency of charge transfer can be tailored by selecting appropriate materials.

In summary, the MoS$_2$ and WS$_2$ QDs were synthesized using chemical assisted exfoliation method. The photo-induced charge transfer between fluorescent R6G  molecules and TMDC QDs of  MoS$_2$ and WS$_2$ was studied using fluorescence spectroscopic techniques. The QDs of MoS$_2$ and WS$_2$ were used as acceptors for charge transfer, which could be further explored to developing QDs-molecules based optoelectronic devices. Maximum charge transfer efficiency was observed in case of MoS$_2$ QDs  which was due  to a wide energy difference between the LUMO of the R6G molecules and the conduction energy level of the MoS$_2$ QDs, it had provided a thermodynamically favourable condition of the charge transfer. 

Authors acknowledge the AIRF, JNU for  TRFS, and TEM characterizations. Authors are also thankful to Dr. Jaydeep Bhattacharya of the School of Biotechnology, JNU for absorption and fluorescence studies.


\begin{thebibliography}{35}%
\makeatletter
\providecommand \@ifxundefined [1]{%
 \@ifx{#1\undefined}
}%
\providecommand \@ifnum [1]{%
 \ifnum #1\expandafter \@firstoftwo
 \else \expandafter \@secondoftwo
 \fi
}%
\providecommand \@ifx [1]{%
 \ifx #1\expandafter \@firstoftwo
 \else \expandafter \@secondoftwo
 \fi
}%
\providecommand \natexlab [1]{#1}%
\providecommand \enquote  [1]{``#1''}%
\providecommand \bibnamefont  [1]{#1}%
\providecommand \bibfnamefont [1]{#1}%
\providecommand \citenamefont [1]{#1}%
\providecommand \href@noop [0]{\@secondoftwo}%
\providecommand \href [0]{\begingroup \@sanitize@url \@href}%
\providecommand \@href[1]{\@@startlink{#1}\@@href}%
\providecommand \@@href[1]{\endgroup#1\@@endlink}%
\providecommand \@sanitize@url [0]{\catcode `\\12\catcode `\$12\catcode `\&12\catcode `\#12\catcode `\^12\catcode `\_12\catcode `\%12\relax}%
\providecommand \@@startlink[1]{}%
\providecommand \@@endlink[0]{}%
\providecommand \url  [0]{\begingroup\@sanitize@url \@url }%
\providecommand \@url [1]{\endgroup\@href {#1}{\urlprefix }}%
\providecommand \urlprefix  [0]{URL }%
\providecommand \Eprint [0]{\href }%
\providecommand \doibase [0]{http://dx.doi.org/}%
\providecommand \selectlanguage [0]{\@gobble}%
\providecommand \bibinfo  [0]{\@secondoftwo}%
\providecommand \bibfield  [0]{\@secondoftwo}%
\providecommand \translation [1]{[#1]}%
\providecommand \BibitemOpen [0]{}%
\providecommand \bibitemStop [0]{}%
\providecommand \bibitemNoStop [0]{.\EOS\space}%
\providecommand \EOS [0]{\spacefactor3000\relax}%
\providecommand \BibitemShut  [1]{\csname bibitem#1\endcsname}%
\let\auto@bib@innerbib\@empty
\bibitem [{\citenamefont {Xu}\ \emph {et~al.}(2021)\citenamefont {Xu}, \citenamefont {Gao}, \citenamefont {Wang}, \citenamefont {Wang}, \citenamefont {Liu},\ and\ \citenamefont {Wang}}]{xu2021quantum}%
  \BibitemOpen
  \bibfield  {author} {\bibinfo {author} {\bibfnamefont {Q.}~\bibnamefont {Xu}}, \bibinfo {author} {\bibfnamefont {J.}~\bibnamefont {Gao}}, \bibinfo {author} {\bibfnamefont {S.}~\bibnamefont {Wang}}, \bibinfo {author} {\bibfnamefont {Y.}~\bibnamefont {Wang}}, \bibinfo {author} {\bibfnamefont {D.}~\bibnamefont {Liu}}, \ and\ \bibinfo {author} {\bibfnamefont {J.}~\bibnamefont {Wang}},\ }\bibfield  {title} {\enquote {\bibinfo {title} {Quantum dots in cell imaging and their safety issues},}\ }\href@noop {} {\bibfield  {journal} {\bibinfo  {journal} {Journal of Materials Chemistry B}\ }\textbf {\bibinfo {volume} {9}},\ \bibinfo {pages} {5765--5779} (\bibinfo {year} {2021})}\BibitemShut {NoStop}%
\bibitem [{\citenamefont {Li}\ \emph {et~al.}(2024)\citenamefont {Li}, \citenamefont {Yan}, \citenamefont {Zhao}, \citenamefont {Ma}, \citenamefont {Zhang}, \citenamefont {Chen}, \citenamefont {Shen}, \citenamefont {Khalaf}, \citenamefont {Zhang}, \citenamefont {Chen} \emph {et~al.}}]{li2024synergistic}%
  \BibitemOpen
  \bibfield  {author} {\bibinfo {author} {\bibfnamefont {M.}~\bibnamefont {Li}}, \bibinfo {author} {\bibfnamefont {J.}~\bibnamefont {Yan}}, \bibinfo {author} {\bibfnamefont {X.}~\bibnamefont {Zhao}}, \bibinfo {author} {\bibfnamefont {T.}~\bibnamefont {Ma}}, \bibinfo {author} {\bibfnamefont {A.}~\bibnamefont {Zhang}}, \bibinfo {author} {\bibfnamefont {S.}~\bibnamefont {Chen}}, \bibinfo {author} {\bibfnamefont {G.}~\bibnamefont {Shen}}, \bibinfo {author} {\bibfnamefont {G.~M.~G.}\ \bibnamefont {Khalaf}}, \bibinfo {author} {\bibfnamefont {J.}~\bibnamefont {Zhang}}, \bibinfo {author} {\bibfnamefont {C.}~\bibnamefont {Chen}},  \emph {et~al.},\ }\bibfield  {title} {\enquote {\bibinfo {title} {Synergistic enhancement of efficient perovskite/quantum dot tandem solar cells based on transparent electrode and band alignment engineering},}\ }\href@noop {} {\bibfield  {journal} {\bibinfo  {journal} {Advanced Energy Materials}\ ,\ \bibinfo {pages} {2400219}} (\bibinfo {year} {2024})}\BibitemShut {NoStop}%
\bibitem [{\citenamefont {Wang}\ \emph {et~al.}(2020)\citenamefont {Wang}, \citenamefont {Li}, \citenamefont {Yin}, \citenamefont {Liu}, \citenamefont {Guo}, \citenamefont {Lai}, \citenamefont {Han}, \citenamefont {Li}, \citenamefont {Li}, \citenamefont {Zhang} \emph {et~al.}}]{wang2020full}%
  \BibitemOpen
  \bibfield  {author} {\bibinfo {author} {\bibfnamefont {L.}~\bibnamefont {Wang}}, \bibinfo {author} {\bibfnamefont {W.}~\bibnamefont {Li}}, \bibinfo {author} {\bibfnamefont {L.}~\bibnamefont {Yin}}, \bibinfo {author} {\bibfnamefont {Y.}~\bibnamefont {Liu}}, \bibinfo {author} {\bibfnamefont {H.}~\bibnamefont {Guo}}, \bibinfo {author} {\bibfnamefont {J.}~\bibnamefont {Lai}}, \bibinfo {author} {\bibfnamefont {Y.}~\bibnamefont {Han}}, \bibinfo {author} {\bibfnamefont {G.}~\bibnamefont {Li}}, \bibinfo {author} {\bibfnamefont {M.}~\bibnamefont {Li}}, \bibinfo {author} {\bibfnamefont {J.}~\bibnamefont {Zhang}},  \emph {et~al.},\ }\bibfield  {title} {\enquote {\bibinfo {title} {Full-color fluorescent carbon quantum dots},}\ }\href@noop {} {\bibfield  {journal} {\bibinfo  {journal} {Science advances}\ }\textbf {\bibinfo {volume} {6}},\ \bibinfo {pages} {eabb6772} (\bibinfo {year} {2020})}\BibitemShut {NoStop}%
\bibitem [{\citenamefont {Sadeghi}\ \emph {et~al.}(2019)\citenamefont {Sadeghi}, \citenamefont {Khabbaz~Abkenar}, \citenamefont {Ow-Yang},\ and\ \citenamefont {Nizamoglu}}]{sadeghi2019efficient}%
  \BibitemOpen
  \bibfield  {author} {\bibinfo {author} {\bibfnamefont {S.}~\bibnamefont {Sadeghi}}, \bibinfo {author} {\bibfnamefont {S.}~\bibnamefont {Khabbaz~Abkenar}}, \bibinfo {author} {\bibfnamefont {C.~W.}\ \bibnamefont {Ow-Yang}}, \ and\ \bibinfo {author} {\bibfnamefont {S.}~\bibnamefont {Nizamoglu}},\ }\bibfield  {title} {\enquote {\bibinfo {title} {Efficient white leds using liquid-state magic-sized cdse quantum dots},}\ }\href@noop {} {\bibfield  {journal} {\bibinfo  {journal} {Scientific reports}\ }\textbf {\bibinfo {volume} {9}},\ \bibinfo {pages} {1--9} (\bibinfo {year} {2019})}\BibitemShut {NoStop}%
\bibitem [{\citenamefont {Li}\ \emph {et~al.}(2016)\citenamefont {Li}, \citenamefont {Wu}, \citenamefont {Zhang}, \citenamefont {Cai}, \citenamefont {Gu}, \citenamefont {Song},\ and\ \citenamefont {Zeng}}]{li2016cspbx3}%
  \BibitemOpen
  \bibfield  {author} {\bibinfo {author} {\bibfnamefont {X.}~\bibnamefont {Li}}, \bibinfo {author} {\bibfnamefont {Y.}~\bibnamefont {Wu}}, \bibinfo {author} {\bibfnamefont {S.}~\bibnamefont {Zhang}}, \bibinfo {author} {\bibfnamefont {B.}~\bibnamefont {Cai}}, \bibinfo {author} {\bibfnamefont {Y.}~\bibnamefont {Gu}}, \bibinfo {author} {\bibfnamefont {J.}~\bibnamefont {Song}}, \ and\ \bibinfo {author} {\bibfnamefont {H.}~\bibnamefont {Zeng}},\ }\bibfield  {title} {\enquote {\bibinfo {title} {Cspbx3 quantum dots for lighting and displays: room-temperature synthesis, photoluminescence superiorities, underlying origins and white light-emitting diodes},}\ }\href@noop {} {\bibfield  {journal} {\bibinfo  {journal} {Advanced Functional Materials}\ }\textbf {\bibinfo {volume} {26}},\ \bibinfo {pages} {2435--2445} (\bibinfo {year} {2016})}\BibitemShut {NoStop}%
\bibitem [{\citenamefont {Xue}\ \emph {et~al.}(2019)\citenamefont {Xue}, \citenamefont {Yang}, \citenamefont {Yuan}, \citenamefont {Zhang}, \citenamefont {Gu}, \citenamefont {Xu}, \citenamefont {Ling}, \citenamefont {Wang}, \citenamefont {Li}, \citenamefont {Zhai} \emph {et~al.}}]{xue2019toward}%
  \BibitemOpen
  \bibfield  {author} {\bibinfo {author} {\bibfnamefont {Y.}~\bibnamefont {Xue}}, \bibinfo {author} {\bibfnamefont {F.}~\bibnamefont {Yang}}, \bibinfo {author} {\bibfnamefont {J.}~\bibnamefont {Yuan}}, \bibinfo {author} {\bibfnamefont {Y.}~\bibnamefont {Zhang}}, \bibinfo {author} {\bibfnamefont {M.}~\bibnamefont {Gu}}, \bibinfo {author} {\bibfnamefont {Y.}~\bibnamefont {Xu}}, \bibinfo {author} {\bibfnamefont {X.}~\bibnamefont {Ling}}, \bibinfo {author} {\bibfnamefont {Y.}~\bibnamefont {Wang}}, \bibinfo {author} {\bibfnamefont {F.}~\bibnamefont {Li}}, \bibinfo {author} {\bibfnamefont {T.}~\bibnamefont {Zhai}},  \emph {et~al.},\ }\bibfield  {title} {\enquote {\bibinfo {title} {Toward scalable pbs quantum dot solar cells using a tailored polymeric hole conductor},}\ }\href@noop {} {\bibfield  {journal} {\bibinfo  {journal} {ACS Energy Letters}\ }\textbf {\bibinfo {volume} {4}},\ \bibinfo {pages} {2850--2858} (\bibinfo {year} {2019})}\BibitemShut {NoStop}%
\bibitem [{\citenamefont {Meng}\ \emph {et~al.}(2024)\citenamefont {Meng}, \citenamefont {Xu}, \citenamefont {Zhang},\ and\ \citenamefont {Wang}}]{meng2024colloidal}%
  \BibitemOpen
  \bibfield  {author} {\bibinfo {author} {\bibfnamefont {L.}~\bibnamefont {Meng}}, \bibinfo {author} {\bibfnamefont {Q.}~\bibnamefont {Xu}}, \bibinfo {author} {\bibfnamefont {J.}~\bibnamefont {Zhang}}, \ and\ \bibinfo {author} {\bibfnamefont {X.}~\bibnamefont {Wang}},\ }\bibfield  {title} {\enquote {\bibinfo {title} {Colloidal quantum dot materials for next-generation near-infrared optoelectronics},}\ }\href@noop {} {\bibfield  {journal} {\bibinfo  {journal} {Chemical Communications}\ } (\bibinfo {year} {2024})}\BibitemShut {NoStop}%
\bibitem [{\citenamefont {Li}\ \emph {et~al.}(2015)\citenamefont {Li}, \citenamefont {Rui}, \citenamefont {Song}, \citenamefont {Shen},\ and\ \citenamefont {Zeng}}]{li2015carbon}%
  \BibitemOpen
  \bibfield  {author} {\bibinfo {author} {\bibfnamefont {X.}~\bibnamefont {Li}}, \bibinfo {author} {\bibfnamefont {M.}~\bibnamefont {Rui}}, \bibinfo {author} {\bibfnamefont {J.}~\bibnamefont {Song}}, \bibinfo {author} {\bibfnamefont {Z.}~\bibnamefont {Shen}}, \ and\ \bibinfo {author} {\bibfnamefont {H.}~\bibnamefont {Zeng}},\ }\bibfield  {title} {\enquote {\bibinfo {title} {Carbon and graphene quantum dots for optoelectronic and energy devices: a review},}\ }\href@noop {} {\bibfield  {journal} {\bibinfo  {journal} {Advanced Functional Materials}\ }\textbf {\bibinfo {volume} {25}},\ \bibinfo {pages} {4929--4947} (\bibinfo {year} {2015})}\BibitemShut {NoStop}%
\bibitem [{\citenamefont {Litvin}\ \emph {et~al.}(2017)\citenamefont {Litvin}, \citenamefont {Martynenko}, \citenamefont {Purcell-Milton}, \citenamefont {Baranov}, \citenamefont {Fedorov},\ and\ \citenamefont {Gun'Ko}}]{litvin2017colloidal}%
  \BibitemOpen
  \bibfield  {author} {\bibinfo {author} {\bibfnamefont {A.}~\bibnamefont {Litvin}}, \bibinfo {author} {\bibfnamefont {I.}~\bibnamefont {Martynenko}}, \bibinfo {author} {\bibfnamefont {F.}~\bibnamefont {Purcell-Milton}}, \bibinfo {author} {\bibfnamefont {A.}~\bibnamefont {Baranov}}, \bibinfo {author} {\bibfnamefont {A.}~\bibnamefont {Fedorov}}, \ and\ \bibinfo {author} {\bibfnamefont {Y.}~\bibnamefont {Gun'Ko}},\ }\bibfield  {title} {\enquote {\bibinfo {title} {Colloidal quantum dots for optoelectronics},}\ }\href@noop {} {\bibfield  {journal} {\bibinfo  {journal} {Journal of Materials Chemistry A}\ }\textbf {\bibinfo {volume} {5}},\ \bibinfo {pages} {13252--13275} (\bibinfo {year} {2017})}\BibitemShut {NoStop}%
\bibitem [{\citenamefont {Algar}\ and\ \citenamefont {Krull}(2008)}]{algar2008quantum}%
  \BibitemOpen
  \bibfield  {author} {\bibinfo {author} {\bibfnamefont {W.~R.}\ \bibnamefont {Algar}}\ and\ \bibinfo {author} {\bibfnamefont {U.~J.}\ \bibnamefont {Krull}},\ }\bibfield  {title} {\enquote {\bibinfo {title} {Quantum dots as donors in fluorescence resonance energy transfer for the bioanalysis of nucleic acids, proteins, and other biological molecules},}\ }\href@noop {} {\bibfield  {journal} {\bibinfo  {journal} {Analytical and bioanalytical chemistry}\ }\textbf {\bibinfo {volume} {391}},\ \bibinfo {pages} {1609--1618} (\bibinfo {year} {2008})}\BibitemShut {NoStop}%
\bibitem [{\citenamefont {Wang}\ \emph {et~al.}(2021)\citenamefont {Wang}, \citenamefont {Chen}, \citenamefont {Liu}, \citenamefont {Liu},\ and\ \citenamefont {Jiang}}]{wang2021thermodynamic}%
  \BibitemOpen
  \bibfield  {author} {\bibinfo {author} {\bibfnamefont {Q.}~\bibnamefont {Wang}}, \bibinfo {author} {\bibfnamefont {W.-Q.}\ \bibnamefont {Chen}}, \bibinfo {author} {\bibfnamefont {X.-Y.}\ \bibnamefont {Liu}}, \bibinfo {author} {\bibfnamefont {Y.}~\bibnamefont {Liu}}, \ and\ \bibinfo {author} {\bibfnamefont {F.-L.}\ \bibnamefont {Jiang}},\ }\bibfield  {title} {\enquote {\bibinfo {title} {Thermodynamic implications and time evolution of the interactions of near-infrared pbs quantum dots with human serum albumin},}\ }\href@noop {} {\bibfield  {journal} {\bibinfo  {journal} {ACS omega}\ }\textbf {\bibinfo {volume} {6}},\ \bibinfo {pages} {5569--5581} (\bibinfo {year} {2021})}\BibitemShut {NoStop}%
\bibitem [{\citenamefont {Chakraborti}\ \emph {et~al.}(2016)\citenamefont {Chakraborti}, \citenamefont {Patel}, \citenamefont {Kanaujia}, \citenamefont {Nath}, \citenamefont {Prakash},\ and\ \citenamefont {Sanyal}}]{chakraborti2016resonance}%
  \BibitemOpen
  \bibfield  {author} {\bibinfo {author} {\bibfnamefont {A.}~\bibnamefont {Chakraborti}}, \bibinfo {author} {\bibfnamefont {A.~S.}\ \bibnamefont {Patel}}, \bibinfo {author} {\bibfnamefont {P.~K.}\ \bibnamefont {Kanaujia}}, \bibinfo {author} {\bibfnamefont {P.}~\bibnamefont {Nath}}, \bibinfo {author} {\bibfnamefont {G.~V.}\ \bibnamefont {Prakash}}, \ and\ \bibinfo {author} {\bibfnamefont {D.}~\bibnamefont {Sanyal}},\ }\bibfield  {title} {\enquote {\bibinfo {title} {Resonance raman scattering and ab initio calculation of electron energy loss spectra of mos2 nanosheets},}\ }\href@noop {} {\bibfield  {journal} {\bibinfo  {journal} {Physics Letters A}\ }\textbf {\bibinfo {volume} {380}},\ \bibinfo {pages} {4057--4061} (\bibinfo {year} {2016})}\BibitemShut {NoStop}%
\bibitem [{\citenamefont {Zhu}, \citenamefont {Zhang},\ and\ \citenamefont {Xia}(2018)}]{zhu2018planar}%
  \BibitemOpen
  \bibfield  {author} {\bibinfo {author} {\bibfnamefont {H.}~\bibnamefont {Zhu}}, \bibinfo {author} {\bibfnamefont {H.}~\bibnamefont {Zhang}}, \ and\ \bibinfo {author} {\bibfnamefont {Y.}~\bibnamefont {Xia}},\ }\bibfield  {title} {\enquote {\bibinfo {title} {Planar is better: monodisperse three-layered mos2 quantum dots as fluorescent reporters for 2, 4, 6-trinitrotoluene sensing in environmental water and luggage cases},}\ }\href@noop {} {\bibfield  {journal} {\bibinfo  {journal} {Analytical chemistry}\ }\textbf {\bibinfo {volume} {90}},\ \bibinfo {pages} {3942--3949} (\bibinfo {year} {2018})}\BibitemShut {NoStop}%
\bibitem [{\citenamefont {Lima}\ \emph {et~al.}(2022)\citenamefont {Lima}, \citenamefont {Oliveira}, \citenamefont {Silva}, \citenamefont {Cabral~Filho}, \citenamefont {Juul-Madsen}, \citenamefont {Moura}, \citenamefont {Vorup-Jensen},\ and\ \citenamefont {Fontes}}]{lima2022mannose}%
  \BibitemOpen
  \bibfield  {author} {\bibinfo {author} {\bibfnamefont {C.~N.}\ \bibnamefont {Lima}}, \bibinfo {author} {\bibfnamefont {W.~F.}\ \bibnamefont {Oliveira}}, \bibinfo {author} {\bibfnamefont {P.~M.}\ \bibnamefont {Silva}}, \bibinfo {author} {\bibfnamefont {P.~E.}\ \bibnamefont {Cabral~Filho}}, \bibinfo {author} {\bibfnamefont {K.}~\bibnamefont {Juul-Madsen}}, \bibinfo {author} {\bibfnamefont {P.}~\bibnamefont {Moura}}, \bibinfo {author} {\bibfnamefont {T.}~\bibnamefont {Vorup-Jensen}}, \ and\ \bibinfo {author} {\bibfnamefont {A.}~\bibnamefont {Fontes}},\ }\bibfield  {title} {\enquote {\bibinfo {title} {Mannose-binding lectin conjugated to quantum dots as fluorescent nanotools for carbohydrate tracing},}\ }\href@noop {} {\bibfield  {journal} {\bibinfo  {journal} {Methods and Applications in Fluorescence}\ }\textbf {\bibinfo {volume} {10}},\ \bibinfo {pages} {025002} (\bibinfo {year} {2022})}\BibitemShut {NoStop}%
\bibitem [{\citenamefont {Wu}\ \emph {et~al.}(2020)\citenamefont {Wu}, \citenamefont {Liu}, \citenamefont {Zhang}, \citenamefont {Wang},\ and\ \citenamefont {Sun}}]{wu2020development}%
  \BibitemOpen
  \bibfield  {author} {\bibinfo {author} {\bibfnamefont {Z.}~\bibnamefont {Wu}}, \bibinfo {author} {\bibfnamefont {P.}~\bibnamefont {Liu}}, \bibinfo {author} {\bibfnamefont {W.}~\bibnamefont {Zhang}}, \bibinfo {author} {\bibfnamefont {K.}~\bibnamefont {Wang}}, \ and\ \bibinfo {author} {\bibfnamefont {X.~W.}\ \bibnamefont {Sun}},\ }\bibfield  {title} {\enquote {\bibinfo {title} {Development of inp quantum dot-based light-emitting diodes},}\ }\href@noop {} {\bibfield  {journal} {\bibinfo  {journal} {ACS Energy Letters}\ }\textbf {\bibinfo {volume} {5}},\ \bibinfo {pages} {1095--1106} (\bibinfo {year} {2020})}\BibitemShut {NoStop}%
\bibitem [{\citenamefont {Rani}\ \emph {et~al.}(2020)\citenamefont {Rani}, \citenamefont {Singh}, \citenamefont {Patel}, \citenamefont {Chakraborti}, \citenamefont {Kumar}, \citenamefont {Ghosh},\ and\ \citenamefont {Sharma}}]{rani2020visible}%
  \BibitemOpen
  \bibfield  {author} {\bibinfo {author} {\bibfnamefont {A.}~\bibnamefont {Rani}}, \bibinfo {author} {\bibfnamefont {K.}~\bibnamefont {Singh}}, \bibinfo {author} {\bibfnamefont {A.~S.}\ \bibnamefont {Patel}}, \bibinfo {author} {\bibfnamefont {A.}~\bibnamefont {Chakraborti}}, \bibinfo {author} {\bibfnamefont {S.}~\bibnamefont {Kumar}}, \bibinfo {author} {\bibfnamefont {K.}~\bibnamefont {Ghosh}}, \ and\ \bibinfo {author} {\bibfnamefont {P.}~\bibnamefont {Sharma}},\ }\bibfield  {title} {\enquote {\bibinfo {title} {Visible light driven photocatalysis of organic dyes using sno2 decorated mos2 nanocomposites},}\ }\href@noop {} {\bibfield  {journal} {\bibinfo  {journal} {Chemical Physics Letters}\ }\textbf {\bibinfo {volume} {738}},\ \bibinfo {pages} {136874} (\bibinfo {year} {2020})}\BibitemShut {NoStop}%
\bibitem [{\citenamefont {Kharangarh}\ \emph {et~al.}(2023)\citenamefont {Kharangarh}, \citenamefont {Ravindra}, \citenamefont {Singh},\ and\ \citenamefont {Umapathy}}]{kharangarh2023synthesis}%
  \BibitemOpen
  \bibfield  {author} {\bibinfo {author} {\bibfnamefont {P.~R.}\ \bibnamefont {Kharangarh}}, \bibinfo {author} {\bibfnamefont {N.~M.}\ \bibnamefont {Ravindra}}, \bibinfo {author} {\bibfnamefont {G.}~\bibnamefont {Singh}}, \ and\ \bibinfo {author} {\bibfnamefont {S.}~\bibnamefont {Umapathy}},\ }\bibfield  {title} {\enquote {\bibinfo {title} {Synthesis of luminescent graphene quantum dots from biomass waste materials for energy-related applications—an overview},}\ }\href@noop {} {\bibfield  {journal} {\bibinfo  {journal} {Energy Storage}\ }\textbf {\bibinfo {volume} {5}},\ \bibinfo {pages} {e390} (\bibinfo {year} {2023})}\BibitemShut {NoStop}%
\bibitem [{\citenamefont {Sariga}\ \emph {et~al.}(2023)\citenamefont {Sariga}, \citenamefont {Babu}, \citenamefont {Kumar}, \citenamefont {Rajeev}, \citenamefont {Thadathil},\ and\ \citenamefont {Varghese}}]{sariga2023new}%
  \BibitemOpen
  \bibfield  {author} {\bibinfo {author} {\bibnamefont {Sariga}}, \bibinfo {author} {\bibfnamefont {A.~M.}\ \bibnamefont {Babu}}, \bibinfo {author} {\bibfnamefont {S.}~\bibnamefont {Kumar}}, \bibinfo {author} {\bibfnamefont {R.}~\bibnamefont {Rajeev}}, \bibinfo {author} {\bibfnamefont {D.~A.}\ \bibnamefont {Thadathil}}, \ and\ \bibinfo {author} {\bibfnamefont {A.}~\bibnamefont {Varghese}},\ }\bibfield  {title} {\enquote {\bibinfo {title} {New horizons in the synthesis, properties, and applications of mxene quantum dots},}\ }\href@noop {} {\bibfield  {journal} {\bibinfo  {journal} {Advanced Materials Interfaces}\ }\textbf {\bibinfo {volume} {10}},\ \bibinfo {pages} {2202139} (\bibinfo {year} {2023})}\BibitemShut {NoStop}%
\bibitem [{\citenamefont {Gan}\ \emph {et~al.}(2015)\citenamefont {Gan}, \citenamefont {Liu}, \citenamefont {Wu}, \citenamefont {Hao}, \citenamefont {Shan}, \citenamefont {Wu},\ and\ \citenamefont {Chu}}]{gan2015quantum}%
  \BibitemOpen
  \bibfield  {author} {\bibinfo {author} {\bibfnamefont {Z.}~\bibnamefont {Gan}}, \bibinfo {author} {\bibfnamefont {L.}~\bibnamefont {Liu}}, \bibinfo {author} {\bibfnamefont {H.}~\bibnamefont {Wu}}, \bibinfo {author} {\bibfnamefont {Y.}~\bibnamefont {Hao}}, \bibinfo {author} {\bibfnamefont {Y.}~\bibnamefont {Shan}}, \bibinfo {author} {\bibfnamefont {X.}~\bibnamefont {Wu}}, \ and\ \bibinfo {author} {\bibfnamefont {P.~K.}\ \bibnamefont {Chu}},\ }\bibfield  {title} {\enquote {\bibinfo {title} {Quantum confinement effects across two-dimensional planes in mos2 quantum dots},}\ }\href@noop {} {\bibfield  {journal} {\bibinfo  {journal} {Applied Physics Letters}\ }\textbf {\bibinfo {volume} {106}},\ \bibinfo {pages} {233113} (\bibinfo {year} {2015})}\BibitemShut {NoStop}%
\bibitem [{\citenamefont {Guo}\ \emph {et~al.}(2019)\citenamefont {Guo}, \citenamefont {Xing}, \citenamefont {Zhao}, \citenamefont {Li}, \citenamefont {Yang},\ and\ \citenamefont {Zhou}}]{guo2019ws2}%
  \BibitemOpen
  \bibfield  {author} {\bibinfo {author} {\bibfnamefont {M.}~\bibnamefont {Guo}}, \bibinfo {author} {\bibfnamefont {Z.}~\bibnamefont {Xing}}, \bibinfo {author} {\bibfnamefont {T.}~\bibnamefont {Zhao}}, \bibinfo {author} {\bibfnamefont {Z.}~\bibnamefont {Li}}, \bibinfo {author} {\bibfnamefont {S.}~\bibnamefont {Yang}}, \ and\ \bibinfo {author} {\bibfnamefont {W.}~\bibnamefont {Zhou}},\ }\bibfield  {title} {\enquote {\bibinfo {title} {Ws2 quantum dots/mos2@ wo3-x core-shell hierarchical dual z-scheme tandem heterojunctions with wide-spectrum response and enhanced photocatalytic performance},}\ }\href@noop {} {\bibfield  {journal} {\bibinfo  {journal} {Applied Catalysis B: Environmental}\ }\textbf {\bibinfo {volume} {257}},\ \bibinfo {pages} {117913} (\bibinfo {year} {2019})}\BibitemShut {NoStop}%
\bibitem [{\citenamefont {Zhu}\ \emph {et~al.}(2016)\citenamefont {Zhu}, \citenamefont {Yang}, \citenamefont {Wu},\ and\ \citenamefont {Lian}}]{zhu2016charge}%
  \BibitemOpen
  \bibfield  {author} {\bibinfo {author} {\bibfnamefont {H.}~\bibnamefont {Zhu}}, \bibinfo {author} {\bibfnamefont {Y.}~\bibnamefont {Yang}}, \bibinfo {author} {\bibfnamefont {K.}~\bibnamefont {Wu}}, \ and\ \bibinfo {author} {\bibfnamefont {T.}~\bibnamefont {Lian}},\ }\bibfield  {title} {\enquote {\bibinfo {title} {Charge transfer dynamics from photoexcited semiconductor quantum dots},}\ }\href@noop {} {\bibfield  {journal} {\bibinfo  {journal} {Annual review of physical chemistry}\ }\textbf {\bibinfo {volume} {67}},\ \bibinfo {pages} {259--281} (\bibinfo {year} {2016})}\BibitemShut {NoStop}%
\bibitem [{\citenamefont {Yan}\ \emph {et~al.}(2016)\citenamefont {Yan}, \citenamefont {Zhang}, \citenamefont {Gu}, \citenamefont {Ding}, \citenamefont {Li},\ and\ \citenamefont {Xian}}]{yan2016facile}%
  \BibitemOpen
  \bibfield  {author} {\bibinfo {author} {\bibfnamefont {Y.}~\bibnamefont {Yan}}, \bibinfo {author} {\bibfnamefont {C.}~\bibnamefont {Zhang}}, \bibinfo {author} {\bibfnamefont {W.}~\bibnamefont {Gu}}, \bibinfo {author} {\bibfnamefont {C.}~\bibnamefont {Ding}}, \bibinfo {author} {\bibfnamefont {X.}~\bibnamefont {Li}}, \ and\ \bibinfo {author} {\bibfnamefont {Y.}~\bibnamefont {Xian}},\ }\bibfield  {title} {\enquote {\bibinfo {title} {Facile synthesis of water-soluble ws2 quantum dots for turn-on fluorescent measurement of lipoic acid},}\ }\href@noop {} {\bibfield  {journal} {\bibinfo  {journal} {The Journal of Physical Chemistry C}\ }\textbf {\bibinfo {volume} {120}},\ \bibinfo {pages} {12170--12177} (\bibinfo {year} {2016})}\BibitemShut {NoStop}%
\bibitem [{\citenamefont {Ghorai}\ \emph {et~al.}(2017)\citenamefont {Ghorai}, \citenamefont {Bayan}, \citenamefont {Gogurla}, \citenamefont {Midya},\ and\ \citenamefont {Ray}}]{ghorai2017highly}%
  \BibitemOpen
  \bibfield  {author} {\bibinfo {author} {\bibfnamefont {A.}~\bibnamefont {Ghorai}}, \bibinfo {author} {\bibfnamefont {S.}~\bibnamefont {Bayan}}, \bibinfo {author} {\bibfnamefont {N.}~\bibnamefont {Gogurla}}, \bibinfo {author} {\bibfnamefont {A.}~\bibnamefont {Midya}}, \ and\ \bibinfo {author} {\bibfnamefont {S.~K.}\ \bibnamefont {Ray}},\ }\bibfield  {title} {\enquote {\bibinfo {title} {Highly luminescent ws2 quantum dots/zno heterojunctions for light emitting devices},}\ }\href@noop {} {\bibfield  {journal} {\bibinfo  {journal} {ACS applied materials \& interfaces}\ }\textbf {\bibinfo {volume} {9}},\ \bibinfo {pages} {558--565} (\bibinfo {year} {2017})}\BibitemShut {NoStop}%
\bibitem [{\citenamefont {Singh}\ \emph {et~al.}(2019)\citenamefont {Singh}, \citenamefont {M.~Yadav}, \citenamefont {Mishra}, \citenamefont {Kumar}, \citenamefont {Tiwari}, \citenamefont {Pandey},\ and\ \citenamefont {Srivastava}}]{singh2019ws2}%
  \BibitemOpen
  \bibfield  {author} {\bibinfo {author} {\bibfnamefont {V.~K.}\ \bibnamefont {Singh}}, \bibinfo {author} {\bibfnamefont {S.}~\bibnamefont {M.~Yadav}}, \bibinfo {author} {\bibfnamefont {H.}~\bibnamefont {Mishra}}, \bibinfo {author} {\bibfnamefont {R.}~\bibnamefont {Kumar}}, \bibinfo {author} {\bibfnamefont {R.}~\bibnamefont {Tiwari}}, \bibinfo {author} {\bibfnamefont {A.}~\bibnamefont {Pandey}}, \ and\ \bibinfo {author} {\bibfnamefont {A.}~\bibnamefont {Srivastava}},\ }\bibfield  {title} {\enquote {\bibinfo {title} {Ws2 quantum dot graphene nanocomposite film for uv photodetection},}\ }\href@noop {} {\bibfield  {journal} {\bibinfo  {journal} {ACS Applied Nano Materials}\ }\textbf {\bibinfo {volume} {2}},\ \bibinfo {pages} {3934--3942} (\bibinfo {year} {2019})}\BibitemShut {NoStop}%
\bibitem [{\citenamefont {Guo}\ and\ \citenamefont {Li}(2020)}]{guo2020mos2}%
  \BibitemOpen
  \bibfield  {author} {\bibinfo {author} {\bibfnamefont {Y.}~\bibnamefont {Guo}}\ and\ \bibinfo {author} {\bibfnamefont {J.}~\bibnamefont {Li}},\ }\bibfield  {title} {\enquote {\bibinfo {title} {Mos2 quantum dots: synthesis, properties and biological applications},}\ }\href@noop {} {\bibfield  {journal} {\bibinfo  {journal} {Materials Science and Engineering: C}\ }\textbf {\bibinfo {volume} {109}},\ \bibinfo {pages} {110511} (\bibinfo {year} {2020})}\BibitemShut {NoStop}%
\bibitem [{\citenamefont {Long}\ \emph {et~al.}(2016)\citenamefont {Long}, \citenamefont {Tao}, \citenamefont {Chiu}, \citenamefont {Tang}, \citenamefont {Fung}, \citenamefont {Chai},\ and\ \citenamefont {Tsang}}]{long2016ws2}%
  \BibitemOpen
  \bibfield  {author} {\bibinfo {author} {\bibfnamefont {H.}~\bibnamefont {Long}}, \bibinfo {author} {\bibfnamefont {L.}~\bibnamefont {Tao}}, \bibinfo {author} {\bibfnamefont {C.~P.}\ \bibnamefont {Chiu}}, \bibinfo {author} {\bibfnamefont {C.~Y.}\ \bibnamefont {Tang}}, \bibinfo {author} {\bibfnamefont {K.~H.}\ \bibnamefont {Fung}}, \bibinfo {author} {\bibfnamefont {Y.}~\bibnamefont {Chai}}, \ and\ \bibinfo {author} {\bibfnamefont {Y.~H.}\ \bibnamefont {Tsang}},\ }\bibfield  {title} {\enquote {\bibinfo {title} {The ws2 quantum dot: preparation, characterization and its optical limiting effect in polymethylmethacrylate},}\ }\href@noop {} {\bibfield  {journal} {\bibinfo  {journal} {Nanotechnology}\ }\textbf {\bibinfo {volume} {27}},\ \bibinfo {pages} {414005} (\bibinfo {year} {2016})}\BibitemShut {NoStop}%
\bibitem [{\citenamefont {Coleman}\ \emph {et~al.}(2011)\citenamefont {Coleman}, \citenamefont {Lotya}, \citenamefont {O’Neill}, \citenamefont {Bergin}, \citenamefont {King}, \citenamefont {Khan}, \citenamefont {Young}, \citenamefont {Gaucher}, \citenamefont {De}, \citenamefont {Smith} \emph {et~al.}}]{coleman2011two}%
  \BibitemOpen
  \bibfield  {author} {\bibinfo {author} {\bibfnamefont {J.~N.}\ \bibnamefont {Coleman}}, \bibinfo {author} {\bibfnamefont {M.}~\bibnamefont {Lotya}}, \bibinfo {author} {\bibfnamefont {A.}~\bibnamefont {O’Neill}}, \bibinfo {author} {\bibfnamefont {S.~D.}\ \bibnamefont {Bergin}}, \bibinfo {author} {\bibfnamefont {P.~J.}\ \bibnamefont {King}}, \bibinfo {author} {\bibfnamefont {U.}~\bibnamefont {Khan}}, \bibinfo {author} {\bibfnamefont {K.}~\bibnamefont {Young}}, \bibinfo {author} {\bibfnamefont {A.}~\bibnamefont {Gaucher}}, \bibinfo {author} {\bibfnamefont {S.}~\bibnamefont {De}}, \bibinfo {author} {\bibfnamefont {R.~J.}\ \bibnamefont {Smith}},  \emph {et~al.},\ }\bibfield  {title} {\enquote {\bibinfo {title} {Two-dimensional nanosheets produced by liquid exfoliation of layered materials},}\ }\href@noop {} {\bibfield  {journal} {\bibinfo  {journal} {Science}\ }\textbf {\bibinfo {volume} {331}},\ \bibinfo {pages} {568--571} (\bibinfo {year} {2011})}\BibitemShut {NoStop}%
\bibitem [{\citenamefont {Li}\ \emph {et~al.}(2017)\citenamefont {Li}, \citenamefont {Jiang}, \citenamefont {Li}, \citenamefont {Ran}, \citenamefont {Zuo}, \citenamefont {Wang}, \citenamefont {Qu}, \citenamefont {Zhao}, \citenamefont {Cheng},\ and\ \citenamefont {Lu}}]{li2017preparation}%
  \BibitemOpen
  \bibfield  {author} {\bibinfo {author} {\bibfnamefont {B.}~\bibnamefont {Li}}, \bibinfo {author} {\bibfnamefont {L.}~\bibnamefont {Jiang}}, \bibinfo {author} {\bibfnamefont {X.}~\bibnamefont {Li}}, \bibinfo {author} {\bibfnamefont {P.}~\bibnamefont {Ran}}, \bibinfo {author} {\bibfnamefont {P.}~\bibnamefont {Zuo}}, \bibinfo {author} {\bibfnamefont {A.}~\bibnamefont {Wang}}, \bibinfo {author} {\bibfnamefont {L.}~\bibnamefont {Qu}}, \bibinfo {author} {\bibfnamefont {Y.}~\bibnamefont {Zhao}}, \bibinfo {author} {\bibfnamefont {Z.}~\bibnamefont {Cheng}}, \ and\ \bibinfo {author} {\bibfnamefont {Y.}~\bibnamefont {Lu}},\ }\bibfield  {title} {\enquote {\bibinfo {title} {Preparation of monolayer mos2 quantum dots using temporally shaped femtosecond laser ablation of bulk mos2 targets in water},}\ }\href@noop {} {\bibfield  {journal} {\bibinfo  {journal} {Scientific Reports}\ }\textbf {\bibinfo {volume} {7}},\ \bibinfo {pages} {1--12} (\bibinfo {year} {2017})}\BibitemShut {NoStop}%
\bibitem [{\citenamefont {Liu}\ \emph {et~al.}(2014)\citenamefont {Liu}, \citenamefont {Kim}, \citenamefont {Kim}, \citenamefont {Ye}, \citenamefont {Kim},\ and\ \citenamefont {Lee}}]{liu2014large}%
  \BibitemOpen
  \bibfield  {author} {\bibinfo {author} {\bibfnamefont {N.}~\bibnamefont {Liu}}, \bibinfo {author} {\bibfnamefont {P.}~\bibnamefont {Kim}}, \bibinfo {author} {\bibfnamefont {J.~H.}\ \bibnamefont {Kim}}, \bibinfo {author} {\bibfnamefont {J.~H.}\ \bibnamefont {Ye}}, \bibinfo {author} {\bibfnamefont {S.}~\bibnamefont {Kim}}, \ and\ \bibinfo {author} {\bibfnamefont {C.~J.}\ \bibnamefont {Lee}},\ }\bibfield  {title} {\enquote {\bibinfo {title} {Large-area atomically thin mos2 nanosheets prepared using electrochemical exfoliation},}\ }\href@noop {} {\bibfield  {journal} {\bibinfo  {journal} {ACS nano}\ }\textbf {\bibinfo {volume} {8}},\ \bibinfo {pages} {6902--6910} (\bibinfo {year} {2014})}\BibitemShut {NoStop}%
\bibitem [{\citenamefont {Wu}\ \emph {et~al.}(2005)\citenamefont {Wu}, \citenamefont {Fan}, \citenamefont {Qiu}, \citenamefont {Yang}, \citenamefont {Siu},\ and\ \citenamefont {Chu}}]{wu2005experimental}%
  \BibitemOpen
  \bibfield  {author} {\bibinfo {author} {\bibfnamefont {X.}~\bibnamefont {Wu}}, \bibinfo {author} {\bibfnamefont {J.}~\bibnamefont {Fan}}, \bibinfo {author} {\bibfnamefont {T.}~\bibnamefont {Qiu}}, \bibinfo {author} {\bibfnamefont {X.}~\bibnamefont {Yang}}, \bibinfo {author} {\bibfnamefont {G.}~\bibnamefont {Siu}}, \ and\ \bibinfo {author} {\bibfnamefont {P.~K.}\ \bibnamefont {Chu}},\ }\bibfield  {title} {\enquote {\bibinfo {title} {Experimental evidence for the quantum confinement effect in 3 c-sic nanocrystallites},}\ }\href@noop {} {\bibfield  {journal} {\bibinfo  {journal} {Physical review letters}\ }\textbf {\bibinfo {volume} {94}},\ \bibinfo {pages} {026102} (\bibinfo {year} {2005})}\BibitemShut {NoStop}%
\bibitem [{\citenamefont {Parsapour}\ \emph {et~al.}(1996)\citenamefont {Parsapour}, \citenamefont {Kelley}, \citenamefont {Craft},\ and\ \citenamefont {Wilcoxon}}]{parsapour1996electron}%
  \BibitemOpen
  \bibfield  {author} {\bibinfo {author} {\bibfnamefont {F.}~\bibnamefont {Parsapour}}, \bibinfo {author} {\bibfnamefont {D.}~\bibnamefont {Kelley}}, \bibinfo {author} {\bibfnamefont {S.}~\bibnamefont {Craft}}, \ and\ \bibinfo {author} {\bibfnamefont {J.}~\bibnamefont {Wilcoxon}},\ }\bibfield  {title} {\enquote {\bibinfo {title} {Electron transfer dynamics in mos2 nanoclusters: Normal and inverted behavior},}\ }\href@noop {} {\bibfield  {journal} {\bibinfo  {journal} {The Journal of chemical physics}\ }\textbf {\bibinfo {volume} {104}},\ \bibinfo {pages} {4978--4987} (\bibinfo {year} {1996})}\BibitemShut {NoStop}%
\bibitem [{\citenamefont {Ha}\ \emph {et~al.}(2014)\citenamefont {Ha}, \citenamefont {Han}, \citenamefont {Choi}, \citenamefont {Park},\ and\ \citenamefont {Seo}}]{ha2014dual}%
  \BibitemOpen
  \bibfield  {author} {\bibinfo {author} {\bibfnamefont {H.~D.}\ \bibnamefont {Ha}}, \bibinfo {author} {\bibfnamefont {D.~J.}\ \bibnamefont {Han}}, \bibinfo {author} {\bibfnamefont {J.~S.}\ \bibnamefont {Choi}}, \bibinfo {author} {\bibfnamefont {M.}~\bibnamefont {Park}}, \ and\ \bibinfo {author} {\bibfnamefont {T.~S.}\ \bibnamefont {Seo}},\ }\bibfield  {title} {\enquote {\bibinfo {title} {Dual role of blue luminescent mos2 quantum dots in fluorescence resonance energy transfer phenomenon},}\ }\href@noop {} {\bibfield  {journal} {\bibinfo  {journal} {Small}\ }\textbf {\bibinfo {volume} {10}},\ \bibinfo {pages} {3858--3862} (\bibinfo {year} {2014})}\BibitemShut {NoStop}%
\bibitem [{\citenamefont {Patel}, \citenamefont {Sahoo},\ and\ \citenamefont {Mohanty}(2014)}]{patel2014}%
  \BibitemOpen
  \bibfield  {author} {\bibinfo {author} {\bibfnamefont {A.~S.}\ \bibnamefont {Patel}}, \bibinfo {author} {\bibfnamefont {H.}~\bibnamefont {Sahoo}}, \ and\ \bibinfo {author} {\bibfnamefont {T.}~\bibnamefont {Mohanty}},\ }\href@noop {} {\bibfield  {journal} {\bibinfo  {journal} {Applied Physics Letters}\ }\textbf {\bibinfo {volume} {105}},\ \bibinfo {pages} {063112} (\bibinfo {year} {2014})}\BibitemShut {NoStop}%
\bibitem [{\citenamefont {Li}\ \emph {et~al.}(2010)\citenamefont {Li}, \citenamefont {Xie}, \citenamefont {Hu}, \citenamefont {Shen}, \citenamefont {Zhou}, \citenamefont {Xiang}, \citenamefont {Zhao},\ and\ \citenamefont {Fang}}]{li2010comparison}%
  \BibitemOpen
  \bibfield  {author} {\bibinfo {author} {\bibfnamefont {Y.}~\bibnamefont {Li}}, \bibinfo {author} {\bibfnamefont {W.}~\bibnamefont {Xie}}, \bibinfo {author} {\bibfnamefont {X.}~\bibnamefont {Hu}}, \bibinfo {author} {\bibfnamefont {G.}~\bibnamefont {Shen}}, \bibinfo {author} {\bibfnamefont {X.}~\bibnamefont {Zhou}}, \bibinfo {author} {\bibfnamefont {Y.}~\bibnamefont {Xiang}}, \bibinfo {author} {\bibfnamefont {X.}~\bibnamefont {Zhao}}, \ and\ \bibinfo {author} {\bibfnamefont {P.}~\bibnamefont {Fang}},\ }\bibfield  {title} {\enquote {\bibinfo {title} {Comparison of dye photodegradation and its coupling with light-to-electricity conversion over tio2 and zno},}\ }\href@noop {} {\bibfield  {journal} {\bibinfo  {journal} {Langmuir}\ }\textbf {\bibinfo {volume} {26}},\ \bibinfo {pages} {591--597} (\bibinfo {year} {2010})}\BibitemShut {NoStop}%
\bibitem [{\citenamefont {Mukherjee}\ \emph {et~al.}(2016)\citenamefont {Mukherjee}, \citenamefont {Maiti}, \citenamefont {Katiyar}, \citenamefont {Das},\ and\ \citenamefont {Ray}}]{mukherjee2016novel}%
  \BibitemOpen
  \bibfield  {author} {\bibinfo {author} {\bibfnamefont {S.}~\bibnamefont {Mukherjee}}, \bibinfo {author} {\bibfnamefont {R.}~\bibnamefont {Maiti}}, \bibinfo {author} {\bibfnamefont {A.~K.}\ \bibnamefont {Katiyar}}, \bibinfo {author} {\bibfnamefont {S.}~\bibnamefont {Das}}, \ and\ \bibinfo {author} {\bibfnamefont {S.~K.}\ \bibnamefont {Ray}},\ }\bibfield  {title} {\enquote {\bibinfo {title} {Novel colloidal mos2 quantum dot heterojunctions on silicon platforms for multifunctional optoelectronic devices},}\ }\href@noop {} {\bibfield  {journal} {\bibinfo  {journal} {Scientific reports}\ }\textbf {\bibinfo {volume} {6}},\ \bibinfo {pages} {1--11} (\bibinfo {year} {2016})}\BibitemShut {NoStop}%
\end{thebibliography}
	


%

\end{document}